\documentclass[aps,prb,twocolumn,floats]{revtex4}
\usepackage{graphicx}
\usepackage{subfigure}
\usepackage{epsfig}
\usepackage{dcolumn}
\usepackage{bm}
\usepackage[ansinew]{inputenc}
\usepackage{amsmath}
\usepackage{amsthm}
\usepackage[T1]{fontenc}
\usepackage{amssymb}
\usepackage{amsfonts}
\usepackage[english]{babel}
\usepackage{enumitem}

\newcommand{\ud}{\,\mathrm{d}}
\newcommand{\ds}{\displaystyle}

\begin{document}

\title{Acoustic waves generated by the spin precession}
\author{R. Zarzuela, E. M. Chudnovsky}
\affiliation{Physics Department, Lehman College,
The City University of New York, 250 Bedford Park Boulevard West,
Bronx, NY 10468-1589, U.S.A.}
\date{\today}

\begin{abstract}
We study generation of acoustic waves by a precessing spin of a nanoparticle deposited on the surface of a solid. Our approach elucidates macroscopic dynamics of the Einstein - de Haas effect. It is based upon solution of parameter-free equations of motion mandated by the conservation of the total angular momentum: spin + mechanical. We show that the amplitude of the acoustic waves generated by the spin precession may be sufficient for detection by a tunneling electron microscope. 
\end{abstract}


\maketitle

\section{Introduction}
One hundred years ago Einstein and de Haas had demonstrated experimentally that the change in the magnetic moment of a freely suspended macroscopic body generated its mechanical rotation due to conservation of the total angular momentum \cite{EdH}. (Back then the concept of spin had not existed yet, magnetic moments of atoms were attributed to the atomic orbital currents carrying angular momentum.) Initially the effect has been intensively studied as a tool to measure the gyromagnetic ratio of solids. Subsequent discovery of the electron spin resonance (ESR) and the ferromagnetic resonance (FMR) provided much more accurate determination of the gyromagnetic ratio, greatly diminishing the significance of the mechanical studies based upon the Einstein - de Haas effect. Meanwhile microscopic mechanisms of the transfer of the spin angular momentum to the rotation of the mechanical system as a whole remained poorly understood, resulting in a number of revisions in recent years \cite{chugarsch05prb,Zhang-Niu}.  

The problem somewhat simplifies when the number of mechanical degrees of freedom is finite. This prompted a significant body of work on spins in mechanical resonators inspired by progress in manufacturing micro- and nanoelectromechanical devices (MEMS and NEMS) \cite{chugar14prb}. Einstein - de Haas effect at the nanoscale has been experimentally studied in magnetic microcantilevers \cite{walmorkab06apl,limimtwal14epl} and theoretically explained by the motion of domain walls \cite{jaachugar09prb}. Switching of magnetic moments by mechanical torques has been proposed \cite{Kovalev-PRL2005,cai-PRA14,CJ-JAP2015}.  Mechanical resonators containing single magnetic molecules have been studied \cite{jaachu09prl,jaachugar10epl},  Experiments have progressed to the measurement of the mechanical coupling of a single molecular spin with a carbon nanotube \cite{ganklyrub13NatNano,ganklyrub13acsNano}. 

The aim of this paper is to generalize the angular-momentum-based approach developed for resonators to systems with infinite number of degrees of freedom. As an example of such problem we study generation of acoustic waves by a magnetic nanoparticle on a solid surface in the ESR setup. The precession of the spin results in the time-dependent mechanical torque acting on the nanoparticle. The parameter-free nature of the torque can be traced to the conservation of the total angular momentum (spin + mechanical). The resulting mechanical vibration of the particle on the surface of the solid generates acoustic waves. The effect is proportional to the ratio of the magnetic anisotropy energy and elastic energy, which has relativistic smallness due to spin-orbit origin of the magnetic anisotropy. Nevertheless, the displacement of the surface in the vicinity of the particle is sufficient to be detected by the tunneling electron microscope (TEM). 

The following argument makes the acoustic TEM detection of spin precession possible even in the presence of stronger effects on the tunneling current. The current, being the polar vector that changes sign due to spatial reflection, can only depend on the magnetic moment squared, because the magnetic moment is an axial vector that does not change sign under spatial reflection. Consequently, the direct interaction between the magnetic moment and the tunneling current can only produce oscillations of the current at double the precession frequency. However, generation of the acoustic waves at the precession frequency will make the tunneling current through the oscillating surface to oscillate at the same frequency. Oscillations of the tunneling current at the precession frequency in the vicinity of the magnetic cluster has been detected in the past \cite{manhamdem89prl,manmukram00prb,durwel02apl}. The proposed explanation was based upon oscillation of the tunneling conductance due to the spin-orbit coupling between the magnetic atom and the tip of the TEM \cite{zhubal02prl,balmar02QIP,fraeribal10prb}. 

The paper is structured as follows. The model of a uniaxial magnetic particle coupled to the elastic matrix via the orientation of its easy magnetization axis is described in Section \ref{Model}. Bulk waves generated by the vibration of the nanoparticle due to the precession of its spin induced by the ac magnetic field are studied in Section \ref{Bulk}. The boundary conditions and the surface waves are worked out in Section \ref{Surface}. Results are discussed in Section \ref{Discussion} where numerical estimates are made for an FePt nanoparticle on a gold surface.

\section{Model}\label{Model}
The nanoparticle is described within the model of an uniaxial magnetic anisotropy, that is, its Hamiltonian operator is given by
\begin{equation}
\label{eq0}
\hat{\mathcal{H}}_{\textrm{CF}}=-D\big(\vec{n}\cdot\hat{\vec{S}}\big)^{2},
\end{equation}
where $\vec{n}$ and $\hat{\vec{S}}$ represent the anisotropy axis and the vector of spin operators of the nanoparticle, respectively, and  $D$ is the anisotropy constant. This spin is embedded into a solid (matrix) that can experience elastic deformations in the form of local twists in the spirit of Einstein - de Haas effect. This leads to the time dependence of the orientation of the anisotropy axis. Rotations of the anisotropy axis are parametrized by the twist angle \cite{LL-elasticity}
\begin{equation}
\label{eq1}
\delta\vec{\phi}(\vec{r},t)=\frac{1}{2}\nabla\times\vec{u}(\vec{r},t),
\end{equation}
where $\vec{u}(\vec{r},t)$ represents the displacement field of the crystal lattice surrounding the particle. Hamiltonian \eqref{eq0} is valid in the crystal frame. The corresponding Hamiltonian in the laboratory frame can be obtained by substituting $\vec{n}$ by the corresponding rotated anisotropy axis: the vector of local twists can be expressed as $\delta\vec{\phi}=\theta\hat{k}$, with $\theta=|\delta\vec{\phi}|$, and any rotated vector can be calculated through Rodrigues' rotation formula:
\begin{align}
\vec{v}\mapsto\vec{v}_{\textrm{rot}}&=\vec{v}\cos\theta+(\hat{k}\times\vec{v})\sin\theta+\hat{k}(\hat{k}\cdot\vec{v})(1-\cos\theta),\nonumber\\
&\simeq\vec{v}+\delta\vec{\phi}\times\vec{v}+\frac{1}{2}\delta\vec{\phi}\times(\delta\vec{\phi}\times\vec{v}),
\end{align}
where in the last step we have considered the regime of small deformations ($|\delta\vec{\phi}|\ll1$) and an expansion up to second order in $\delta\vec{\phi}$. Therefore, the following expression for the Hamiltonian operator in the laboratory frame is obtained:
\begin{align}
\hat{\mathcal{H}}_{\textrm{LF}}&
\simeq\hat{\mathcal{H}}_{\textrm{CF}}-D\big\{\vec{n}\cdot\hat{\vec{S}},\vec{n}\times\hat{\vec{S}}\big\}\cdot\delta\vec{\phi}-D\big[\big(\delta\vec{\phi}\times\vec{n}\big)\cdot\hat{\vec{S}}\big]^{2}\nonumber\\
&\quad-\frac{D}{2}\big\{\vec{n}\cdot\hat{\vec{S}},\big(\delta\vec{\phi}\times(\delta\vec{\phi}\times\vec{n})\big)\cdot\hat{\vec{S}}\big\},\\
&=-D\hat{S}_{z}^{2}+D\{\hat{S}_{z},\hat{S}_{y}\}\delta\phi_{x}-D\{\hat{S}_{z},\hat{S}_{x}\}\delta\phi_{y}\nonumber\\
&\quad-\frac{D}{2}\{\hat{S}_{z},\hat{S}_{x}\}\delta\phi_{x}\delta\phi_{z}-\frac{D}{2}\{\hat{S}_{z},\hat{S}_{y}\}\delta\phi_{y}\delta\phi_{z}\nonumber\\
&\quad+D\hat{S}_{z}^{2}\left(\delta\phi_{x}^{2}+\delta\phi_{y}^{2}\right)-D\hat{S}_{x}^{2}\delta\phi_{y}^{2}-D\hat{S}_{y}^{2}\delta\phi_{x}^{2}\nonumber\\
&\quad+D\{\hat{S}_{x},\hat{S}_{y}\}\delta\phi_{x}\delta\phi_{y},\label{eq4}
\end{align}
where in the last step the anisotropy axis has been taken along the $Z$ axis ($\vec{n}=\hat{e}_{z}$) in the crystal frame.

We consider two magnetic fields applied to the nanoparticle: the first one consists of a static field along the anisotropy axis in the crystal frame, whereas the second one is an AC field transversal to it. These fields are given by 
\begin{eqnarray}
\vec{H}_{0}&=&H_{0}\hat{e}_{z},\label{eq6}\\
\vec{H}_{1}&=&h_{x}(t)\hat{e}_{x},\quad h_{x}(t)=\textrm{Re}\left[\mathcal{A}e^{-i\xi t}\right],\nonumber
\end{eqnarray}
which add a Zeeman term to the total Hamiltonian, $\hat{\mathcal{H}}_{\textrm{Z}}=-\gamma\hbar\hat{\vec{S}}\cdot\big(\vec{H}_{0}+\vec{H}_{1}\big)$. We will assume that both $H_{0}$ and $\mathcal{A}$ are spatially homogeneous. From this point forward we also assume that the nanoparticle is located at the origin of coordinates, which translates into the local twists being evaluated at $\vec{r}=\vec{0}$. Therefore, the Hamiltonian operator in the laboratory frame turns out to be
\begin{align}
\label{eq5}
\hat{\mathcal{H}}_{\textrm{LF}}&=-D\hat{S}_{z}^{2}-\gamma\hbar\hat{S}_{z}H_{0}-\gamma\hbar\hat{S}_{x}h_{x}(t)\\
&\quad-D\int_{\mathbf{R}^{3}}\ud^{3}\vec{r}\;\delta^{(3)}(\vec{r})\Big[\{\hat{S}_{z},\hat{S}_{x}\}\delta\phi_{y}(\vec{r},t)\nonumber\\
&\quad-\{\hat{S}_{z},\hat{S}_{y}\}\delta\phi_{x}(\vec{r},t)+\frac{1}{2}\{\hat{S}_{z},\hat{S}_{x}\}\delta\phi_{x}(\vec{r},t)\delta\phi_{z}(\vec{r},t)\nonumber\\
&\quad+\frac{1}{2}\{\hat{S}_{z},\hat{S}_{y}\}\delta\phi_{y}(\vec{r},t)\delta\phi_{z}(\vec{r},t)+\hat{S}_{x}^{2}\delta\phi_{y}^{2}(\vec{r},t)\nonumber\\
&\quad+\hat{S}_{y}^{2}\delta\phi_{x}^{2}(\vec{r},t)-\hat{S}_{z}^{2}\Big(\delta\phi_{x}^{2}(\vec{r},t)+\delta\phi_{y}^{2}(\vec{r},t)\Big)\nonumber\\
&\quad-\{\hat{S}_{x},\hat{S}_{y}\}\delta\phi_{x}(\vec{r},t)\delta\phi_{y}(\vec{r},t)\Big]+\hat{\mathcal{H}}_{\textrm{ph}},\nonumber
\end{align}
where we have included the contribution $\hat{\mathcal{H}}_{\textrm{ph}}$ to the total Hamiltonian from phonons. The classical spin is defined as $\vec{\Sigma}(t)=\textrm{Tr}\big[\hat{\rho}\hat{\vec{S}}\big]$, where $\hat{\rho}$ is the density matrix operator of the system, and the total energy of the system becomes:
\begin{align}
\mathcal{E}&\equiv\textrm{Tr}\big[\hat{\rho}\hat{\mathcal{H}}_{\textrm{LF}}\big]\label{eq10}\\
&=\mathcal{E}_{\textrm{ph}}-D\Sigma_{z}^{2}(t)-\gamma\hbar\Sigma_{z}(t)H_{0}-\gamma\hbar\Sigma_{x}(t)h_{x}(t)\nonumber\\
&\quad-D\int_{\mathbf{R}^{3}}\ud^{3}\vec{r}\;\delta^{(3)}(\vec{r})\Big[2\Sigma_{z}(t)\Sigma_{x}(t)\delta\phi_{y}(\vec{r},t)\nonumber\\
&\quad-2\Sigma_{z}(t)\Sigma_{y}(t)\delta\phi_{x}(\vec{r},t)+\Sigma_{z}(t)\Sigma_{x}(t)\delta\phi_{x}(\vec{r},t)\delta\phi_{z}(\vec{r},t)\nonumber\\
&\quad+\Sigma_{z}(t)\Sigma_{y}(t)\delta\phi_{y}(\vec{r},t)\delta\phi_{z}(\vec{r},t)+\Sigma_{x}^{2}(t)\delta\phi_{y}^{2}(\vec{r},t)\nonumber\\
&\quad+\Sigma_{y}^{2}(t)\delta\phi_{x}^{2}(\vec{r},t)-\Sigma_{z}^{2}(t)\big(\delta\phi_{x}^{2}(\vec{r},t)+\delta\phi_{y}^{2}(\vec{r},t)\big)\nonumber\\
&\quad-2\Sigma_{x}(t)\Sigma_{y}(t)\delta\phi_{x}(\vec{r},t)\delta\phi_{y}(\vec{r},t)\Big].\nonumber
\end{align}

Notice that in this treatment we are assuming that averaging over spin and phonon states can be done independently, a legitimate assumption in the long-wave limit ($|\delta\vec{\phi}|\ll1$). As we are studying an AC-field-driven dynamical system, in the linear regime we have $\Sigma_{x,y}\sim\mathcal{A}$ and $\vec{u}\sim\mathcal{A}$. Therefore, we can linearize (up to first order terms in $\mathcal{A}$) the dynamical equations arising from the total energy $\mathcal{E}$ due to the perturbative nature of this field.

Landau-Lifshitz-Gilbert equation determines the dynamical evolution of the classical spin:
\begin{equation}
\label{eq11}
\hbar\frac{\partial}{\partial t}\vec{\Sigma}=\vec{\Sigma}\times\vec{H}_{\textrm{eff}}-\eta\hbar\frac{\vec{\Sigma}}{|\vec{\Sigma}|}\times\frac{\partial}{\partial t}\vec{\Sigma},
\end{equation}
where $\ds \vec{H}_{\textrm{eff}}=-\delta\mathcal{E}/\delta\vec{\Sigma}$ is the effective field stemming from the total energy of the system and $\eta$ is the dissipation constant. Eq. \eqref{eq10} and linearization leads to the following system of equations: 
\begin{eqnarray}
\hbar\frac{d\Sigma_{x}}{d t}&=&2D\Sigma_{z}\Sigma_{y}+\gamma\hbar H_{0}\Sigma_{y}+2D\Sigma_{z}^{2}\delta\phi_{x}(\vec{r}=\vec{0},t)\nonumber\\
&&-\frac{\eta\hbar}{|\vec{\Sigma}|}\left(\Sigma_{y}\frac{d\Sigma_{z}}{dt}-\Sigma_{z}\frac{\Sigma_{y}}{dt}\right),\label{eq17}\\
\hbar\frac{d\Sigma_{y}}{d t}&=&-2D\Sigma_{z}\Sigma_{x}-\gamma\hbar H_{0}\Sigma_{x}+2D\Sigma_{z}^{2}\delta\phi_{y}(\vec{r}=\vec{0},t)\nonumber\\
&&+\gamma\hbar \Sigma_{z}h_{x}(t)-\frac{\eta\hbar}{|\vec{\Sigma}|}\left(\Sigma_{z}\frac{d\Sigma_{x}}{dt}-\Sigma_{x}\frac{d\Sigma_{z}}{dt}\right),\nonumber\\
\hbar\frac{d\Sigma_{z}}{d t}&=&O(|\mathcal{A}|^{2}).\nonumber
\end{eqnarray}

The third equation leads to the identity $\Sigma_{z}(t)=\Sigma_{z,0}\equiv\,$cte up to second order in $\mathcal{A}$. Furthermore, $|\vec{\Sigma}|\simeq\Sigma_{z,0}$ due to the $x-$ and $y-$components of the classical spin being perturbations. Introducing these dependences into the first two equations, we obtain the linear system of ODEs
\begin{eqnarray}
\hbar\frac{d\Sigma_{x}}{d t}&=&\hbar\omega_{\textrm{FM}}\Sigma_{y}+2D\Sigma_{z,0}^{2}\delta\phi_{x}(\vec{r}=\vec{0},t)+\eta\hbar\frac{d\Sigma_{y}}{dt},\nonumber\\
\hbar\frac{d\Sigma_{y}}{d t}&=&-\hbar\omega_{\textrm{FM}}\Sigma_{x}+2D\Sigma_{z,0}^{2}\delta\phi_{y}(\vec{r}=\vec{0},t)-\eta\hbar\frac{d\Sigma_{x}}{dt}\nonumber\\
&&+\gamma\hbar \Sigma_{z,0}h_{x}(t),\label{eq18}
\end{eqnarray}
where $\ds \omega_{\textrm{FM}}=\frac{2D\Sigma_{z,0}}{\hbar}+\gamma H_{0}$ is the ferromagnetic resonance frequency of the nanoparticle.

The dynamical equation for the displacement field is given by
\begin{equation}
\label{eq13}
\rho\frac{\partial^{2}u_{\alpha}}{\partial t^{2}}=\sum_{\beta}\frac{\partial\sigma_{\alpha\beta}}{\partial x_{\beta}},
\end{equation}
where $\sigma_{\alpha\beta}=\delta\mathcal{E}/\delta e_{\alpha\beta}$ is the stress tensor, $e_{\alpha\beta}=\partial u_{\alpha}/\partial x_{\beta}$ is the strain tensor and $\rho$ is the mass density. Notice that $\delta\vec{\phi}=(1/2)\nabla\times\vec{u}$, from which we straightforwardly deduce the identity $\delta\phi_{\gamma}=(1/2)\epsilon_{\gamma\beta\alpha} e_{\alpha\beta}$. Therefore, the (linear) magnetic contribution to the stress tensor of the lattice is given by
\begin{align}
\label{eq19}
\sigma_{\alpha\beta}^{(a)}=&D\Sigma_{z,0}\Sigma_{y}\delta^{(3)}(\vec{r})\epsilon_{x\beta\alpha}-D\Sigma_{z,0}\Sigma_{x}\delta^{(3)}(\vec{r})\epsilon_{y\beta\alpha}\\
&+D\Sigma_{z,0}^{2}\delta^{(3)}(\vec{r})\left[\delta\phi_{x}(\vec{r},t)\epsilon_{x\beta\alpha}+\delta\phi_{y}(\vec{r},t)\epsilon_{y\beta\alpha}\right].\nonumber
\end{align}

\section{Bulk waves} \label{Bulk}
The dynamics of bulk acoustic waves are described by Eqs. \eqref{eq13}. With account of expression \eqref{eq19} for the magnetic contribution to the stress tensor, we obtain the following dynamical equations for the bulk waves:
\begin{align}
&\frac{\partial^{2}u_{x}}{\partial t^{2}}-c_{t}^{2}\nabla^{2}u_{x}=-\frac{D}{\rho}\Sigma_{z,0}\Sigma_{x}\frac{\partial}{\partial z}\delta^{(3)}(\vec{r})\label{eq17}\\
&\hspace{3.4cm}+\frac{D\Sigma_{z,0}^{2}}{\rho}\frac{\partial}{\partial z}\left[\delta^{(3)}(\vec{r})\delta\phi_{y}(\vec{r},t)\right],\nonumber\\
&\frac{\partial^{2}u_{y}}{\partial t^{2}}-c_{t}^{2}\nabla^{2}u_{y}=-\frac{D}{\rho}\Sigma_{z,0}\Sigma_{y}\frac{\partial}{\partial z}\delta^{(3)}(\vec{r})\nonumber\\
&\hspace{3.4cm}-\frac{D\Sigma_{z,0}^{2}}{\rho}\frac{\partial}{\partial z}\left[\delta^{(3)}(\vec{r})\delta\phi_{x}(\vec{r},t)\right],\nonumber\\
&\frac{\partial^{2}u_{z}}{\partial t^{2}}-c_{t}^{2}\nabla^{2}u_{z}=\frac{D}{\rho}\Sigma_{z,0}\left[\Sigma_{y}\frac{\partial}{\partial y}\delta^{(3)}(\vec{r})+\Sigma_{x}\frac{\partial}{\partial x}\delta^{(3)}(\vec{r})\right]\nonumber\\
&+\frac{D\Sigma_{z,0}^{2}}{\rho}\left(\frac{\partial}{\partial y}\left[\delta^{(3)}(\vec{r})\delta\phi_{x}(\vec{r},t)\right]-\frac{\partial}{\partial x}\left[\delta^{(3)}(\vec{r})\delta\phi_{y}(\vec{r},t)\right]\right),\nonumber
\end{align}
where we have set condition $\nabla\cdot\vec{u}=0$ for the displacement field. This stems from the fact that for nanoparticles, which are more rigid than their environment, only transverse phonons interact with the magnetic degrees of freedom, that is, local twists are transverse in nature. Performing the Fourier transform 
\begin{equation}
f(\vec{r},t)\mapsto \hat{f}(\vec{k},\omega)=\frac{1}{(2\pi)^{2}}\int_{\mathbf{R}^{4}}\ud t\ud^{3}\vec{r}\; f(\vec{r},t)e^{-i\vec{k}\cdot\vec{r}+i\omega t}
\end{equation}
on Eqs. \eqref{eq17} we obtain the algebraic identities
\begin{eqnarray}
\hat{u}_{x}(\vec{k},\omega)&=&\frac{1}{(2\pi)^{3/2}\rho}\frac{i k_{z}}{c_{t}^{2}\vec{k}^{2}-\omega^{2}}\Big[-D\Sigma_{z,0}\hat{\Sigma}_{x}(\omega)\label{eq23}\\
&&\hspace{2cm}+D\Sigma_{z,0}^{2}\mathcal{F}[\delta\phi_{y}(\vec{r}=\vec{0},t)](\omega)\Big]\nonumber\\
\hat{u}_{y}(\vec{k},\omega)&=&-\frac{1}{(2\pi)^{3/2}\rho}\frac{i k_{z}}{c_{t}^{2}\vec{k}^{2}-\omega^{2}}\Big[D\Sigma_{z,0}\hat{\Sigma}_{y}(\omega),\nonumber\\
&&\hspace{2cm}+D\Sigma_{z,0}^{2}\mathcal{F}[\delta\phi_{x}(\vec{r}=\vec{0},t)](\omega)\Big]\nonumber\\
\hat{u}_{z}(\vec{k},\omega)&=&\frac{1}{(2\pi)^{3/2}\rho}\frac{1}{c_{t}^{2}\vec{k}^{2}-\omega^{2}}
\Big[D\Sigma_{z,0}\Big((i k_{x})\hat{\Sigma}_{x}(\omega)+\nonumber\\
&&(i k_{y})\hat{\Sigma}_{y}(\omega)\Big)+D\Sigma_{z,0}^{2}\Big((i k_{y})\mathcal{F}[\delta\phi_{x}(\vec{r}=\vec{0},t)](\omega)\nonumber\\
&&-(i k_{x})\mathcal{F}[\delta\phi_{y}(\vec{r}=\vec{0},t)](\omega)\Big).\nonumber
\end{eqnarray}
With account of the identities \eqref{eq18} we can express the Fourier transforms of the twist angles in terms of $\hat{\vec{\Sigma}}$. Therefore, the above equations become
\begin{eqnarray}
\hat{u}_{x}(\vec{k},\omega)&=&\frac{1}{(2\pi)^{3/2}\rho}\frac{i k_{z}}{c_{t}^{2}\vec{k}^{2}-\omega^{2}}\frac{\hbar}{2}a,\label{eq24}\\
\hat{u}_{y}(\vec{k},\omega)&=&\frac{1}{(2\pi)^{3/2}\rho}\frac{i k_{z}}{c_{t}^{2}\vec{k}^{2}-\omega^{2}}\frac{\hbar}{2}b,\nonumber\\
\hat{u}_{z}(\vec{k},\omega)&=&-\frac{1}{(2\pi)^{3/2}\rho}\frac{\hbar}{2}\frac{(i k_{x})a+(i k_{y})b}{c_{t}^{2}\vec{k}^{2}-\omega^{2}},\nonumber
\end{eqnarray}
with 
\begin{align}
a&=-i\omega\hat{\Sigma}_{y}+\left(\omega_{\textrm{FM}}-\frac{2D\Sigma_{z,0}}{\hbar}-i\omega\eta\right)\hat{\Sigma}_{x}-\gamma\Sigma_{z,0}\hat{h}_{x},\nonumber\\
b&=i\omega\hat{\Sigma}_{x}+\left(\omega_{\textrm{FM}}-\frac{2D\Sigma_{z,0}}{\hbar}-i\omega\eta\right)\hat{\Sigma}_{y}.
\end{align}

Notice that the identities
\begin{align}
\mathcal{F}[\delta\phi_{x}(\vec{0},t)](\omega)&=\int_{\mathbf{R}^{3}}\frac{i\ud^{3}\vec{k}}{2(2\pi)^{3/2}}\left[k_{y}\hat{u}_{z}(\vec{k},\omega)-k_{z}\hat{u}_{y}(\vec{k},\omega)\right],\label{eqFTdeltas}\\
\mathcal{F}[\delta\phi_{y}(\vec{0},t)](\omega)&=\int_{\mathbf{R}^{3}}\frac{i\ud^{3}\vec{k}}{2(2\pi)^{3/2}}\left[k_{z}\hat{u}_{x}(\vec{k},\omega)-k_{x}\hat{u}_{z}(\vec{k},\omega)\right],\nonumber
\end{align}
hold, which combined with Eqs. \eqref{eq23} lead to
\begin{align}
\label{eq28}
\mathcal{F}[\delta\phi_{x}(\vec{r}=\vec{0},t)](\omega)&=\frac{\hbar}{3(2\pi)^{2}\rho c_{t}^{2}}\Xi(\omega)b,\\
\mathcal{F}[\delta\phi_{y}(\vec{r}=\vec{0},t)](\omega)&=-\frac{\hbar}{3(2\pi)^{2}\rho c_{t}^{2}}\Xi(\omega)a.\nonumber
\end{align}
The function $\ds\Xi(\omega)=(2/3)k_{3D}^{3}+i\pi(\omega/c_{t})^{3}$ stems from integration over the 3D wave-vector space of the functions $\ds \frac{k_{z}^{2}+k_{l}^{2}}{c_{t}^{2}\vec{k}^{2}-\omega^{2}},\; l\in\{x,y\}$. The cut-off $\ds k_{3D}=\sqrt{3}\frac{2\pi}{L}$ represents the maximum modulus of a wave vector generated by the precession of an object of size $L$. These integrals can be calculated by means of contour integration in the complex plane, leading to the value $(8\pi/3c_{t}^{2})\Xi(\omega)$. 

In the linear regime the identities
\begin{align}
\hat{\Sigma}_{x}(\omega)=\hat{\chi}_{xx}(\omega)\hat{h}_{x}(\omega),\quad\hat{\Sigma}_{y}(\omega)=\hat{\chi}_{yx}(\omega)\hat{h}_{x}(\omega),
\end{align}
hold. Taking into consideration the functional dependencies \eqref{eq28} of the magnetoelastic coupling terms, the dynamical equations for the spin susceptibility in the Fourier space become:
\begin{equation}
\label{eq29}
\left({\begin{array}{cc}
i\omega(1+\zeta)&M_{12}\\
-M_{12}& 
i\omega(1+\zeta)
  \end{array} } \right)\cdot
  \left(
  \begin{array}{c}
  \hat{\chi}_{xx}\\
  \hat{\chi}_{yx}\\
  \end{array}
  \right)=-\gamma\Sigma_{z,0}
  \left(
  \begin{array}{c}
 0\\
 1+\zeta
  \end{array}
  \right),
\end{equation}
where 
\begin{equation}
\label{zeta}
M_{12}=(\omega_{\textrm{FM}}-i\omega\eta)(1+\zeta)-\frac{2D\Sigma_{z,0}}{\hbar}\zeta,\quad \zeta(\omega)=\frac{2D\Sigma_{z,0}^{2}\Xi(\omega)}{3(2\pi)^{2}\rho c_{t}^{2}}.
\end{equation}

This system of equations can be solved using standard methods, so that we obtain the following expressions for the spin susceptibility 
\begin{equation}
\label{eq30}
  \left(
  \begin{array}{c}
  \hat{\chi}_{xx}\\
  \hat{\chi}_{yx}\\
  \end{array}
  \right)=\frac{\gamma \Sigma_{z,0}(1+\zeta)}{K(\omega)}
  \left(\begin{array}{c}
-M_{12}\\
 i\omega(1+\zeta)\\
  \end{array}
  \right),
\end{equation}
where 
\begin{align}
K(\omega)&=\omega^{2}\left(1+\zeta\right)^{2}-M_{12}^{2}.
\end{align}
We are interested in the regime $|\zeta|\ll1$, within which functions $a$ and $b$ have the following functional dependence:
\begin{align}
\label{eqab}
a&=(\gamma\Sigma_{z,0})\left(\frac{2D\Sigma_{z,0}}{\hbar}\right)(\omega_{\textrm{FM}}-i\omega\eta)\hat{h}_{x}(\omega)/K(\omega),\\
b&=(\gamma \Sigma_{z,0})\left(\frac{2D\Sigma_{z,0}}{\hbar}\right)(-i\omega)\hat{h}_{x}(\omega)/K(\omega)\nonumber
\end{align}

Now we proceed to calculate the time and spatial dependences of the displacement field, that is, the profiles of the acoustic waves generated by the precessional motion of a single spin. Introducing the above functional dependences  for the spin susceptibilities into Eqs. \eqref{eq24}, we obtain:
\begin{align}
u_{x}(\vec{r},t)&=\frac{1}{(2\pi)^{7/2}\rho}\frac{\hbar}{2}\int_{\mathbf{R}}\ud\omega e^{-i\omega t}a\int_{\mathbf{R}^{3}}\ud^{3}\vec{k}\;e^{i\vec{k}\cdot\vec{r}}\frac{i k_{z}}{c_{t}^{2}\vec{k}^{2}-\omega^{2}}\nonumber\\
&=-\frac{i\hbar}{4(2\pi)^{3/2}\rho c_{t}^{2}}\int_{\mathbf{R}}\ud\omega e^{-i\omega t}a\left(\frac{\omega}{c_{t}}\right)^{2}\frac{z}{r}j_{1}[(\omega/c_{t})r],\nonumber
\end{align}
where in the last step we have used the planar wave expansion
\begin{equation}
\label{pwave}
e^{i\vec{k}\cdot\vec{r}}=4\pi\sum_{l\geq0}i^{l}j_{l}[kr]\sum_{m=-l}^{l}Y_{l}^{m\star}(\hat{k})Y_{l}^{m}(\hat{r})
\end{equation}
for arbitrary position vector $\vec{r}$, orthonormality of the spherical harmonics over the 3D sphere and contour integration in the complex plane with respect to the variable $k$ to calculate the integral over the 3D wave-vector space. The spherical Bessel function $\ds j_{1}[x]=\frac{\sin x-x\cos x}{x^{2}}$ determines the spatial dependence of the $x$-component of the displacement field. With account of the identity $\hat{h}_{x}(\omega)=(2\pi)^{1/2}\mathcal{A}\delta^{(1)}(\omega-\xi)$ for a circular AC field, we finally obtain after integration over the frequency domain:
\begin{align}
\label{eq33}
u_{x}(\vec{r},t)=&\frac{\gamma\hbar \Sigma_{z,0}\mathcal{A}}{8\pi\rho c_{t}^{2}}\left(\frac{2D\Sigma_{z,0}}{\hbar}\right)\frac{z}{r}A[\xi,t]\left(\frac{\xi}{c_{t}}\right)^{2}j_{1}[(\xi/c_{t})r],
\end{align}
with $\mathcal{A}\in\mathbf{R}$. We can proceed analogously with the $y$ and $z$-components of the displacement field: 
\begin{align}
\label{eq34}
u_{y}(\vec{r},t)=&\frac{\gamma\hbar \Sigma_{z,0}\mathcal{A}}{8\pi\rho c_{t}^{2}}\left(\frac{2D\Sigma_{z,0}}{\hbar}\right)\frac{z}{r}B[\xi,t]\left(\frac{\xi}{c_{t}}\right)^{2}j_{1}[(\xi/c_{t})r],\nonumber\\
u_{z}(\vec{r},t)=&-\frac{\gamma\hbar \Sigma_{z,0}\mathcal{A}}{8\pi\rho c_{t}^{2}}\left(\frac{2D\Sigma_{z,0}}{\hbar}\right)\left[\frac{x}{r}A[\xi,t]+\frac{y}{r}B[\xi,t]\right]\nonumber\\
&\hspace{1.2cm}\times\left(\frac{\xi}{c_{t}}\right)^{2}j_{1}[(\xi/c_{t})r],
\end{align}
with 
\begin{align}
\label{eq33b}
A[\xi,t]&=\textrm{Re}\left[\frac{\omega_{\textrm{FM}}-i\xi\eta}{iK(\xi)}e^{-i\xi t}\right],\\
 B[\xi,t]&=\textrm{Re}\left[\frac{-\xi}{K(\xi)}e^{-i\xi t}\right],\nonumber
\end{align}
where $K(\omega)=\omega^{2}-(\omega_{\textrm{FM}}-i\omega\eta)^{2}=(\omega^{2}-\omega_{\textrm{FM}}^{2}+\omega^{2}\eta^{2})+2i\omega\omega_{\textrm{FM}}\eta$. At resonance we have  $K(\xi\simeq\omega_{\textrm{FM}})\simeq 2i\omega_{\textrm{FM}}^{2}\eta$, where condition $\eta\ll1$ has been taken into consideration. Therefore, the expressions for the functions $A$ and $B$ close to resonance are:
\begin{align}
\label{eq35}
A[\xi\simeq\omega_{\textrm{FM}},t]&=\frac{1}{2\omega_{\textrm{FM}}\eta}[\eta\sin(\omega_{\textrm{FM}}t)-\cos(\omega_{\textrm{FM}}t)],\nonumber\\
B[\xi\simeq\omega_{\textrm{FM}},t]&=\frac{1}{2\omega_{\textrm{FM}}\eta}\sin(\omega_{\textrm{FM}}t).
\end{align}

\section{Surface waves}\label{Surface}

The geometry of the problem is the following: the surface of the solid is located at the $XY$ plane and its bulk extends over the $Z<0$ space. The mechanical stress tensor is given by: 
\begin{eqnarray}
\label{eq36}
\sigma_{xx}^{m}&=&\frac{E}{(1+\sigma)(1-2\sigma)}\left[(1-\sigma)u_{xx}+\sigma(u_{yy}+u_{zz})\right],\\
\sigma_{yy}^{m}&=&\frac{E}{(1+\sigma)(1-2\sigma)}\left[(1-\sigma)u_{yy}+\sigma(u_{xx}+u_{zz})\right],\nonumber\\
\sigma_{zz}^{m}&=&\frac{E}{(1+\sigma)(1-2\sigma)}\left[(1-\sigma)u_{zz}+\sigma(u_{xx}+u_{yy})\right],\nonumber\\
\sigma_{xy}^{m}&=&\frac{E}{(1+\sigma)}u_{xy},\hspace{0.1cm}\sigma_{xz}=\frac{E}{(1+\sigma)}u_{xz},\hspace{0.1cm}\sigma_{yz}=\frac{E}{(1+\sigma)}u_{yz},\nonumber
\end{eqnarray}
where $u_{\alpha\beta}=\frac{1}{2}(\partial_{\beta}u_{\alpha}+\partial_{\alpha}u_{\beta})$, $E$ is the Young modulus and $\sigma$ is the Poisson ratio. Notice that the transverse and longitudinal speeds of sound are given by $c_{t}^{2}=E/2\rho(1+\sigma)$ and $c_{l}^{2}=E(1-\sigma)/\rho(1+\sigma)(1-2\sigma)$, respectively.

Boundary conditions at the surface are given by $\sigma_{ij}^{m}n_{j}=0$, where $\vec{n}$ is the normal vector to the surface. In our case $n_{j}=\delta_{j,z}$, so that $\sigma_{xz}^{m}=\sigma_{yz}^{m}=\sigma_{zz}^{m}=0$ at $z=0$. The total displacement field is a superposition of transversal and longitudinal modes, with the transversal mode given by the bulk solutions \eqref{eq33} and \eqref{eq34}. The expected functional form for the longitudinal surface wave is given by
\begin{equation}
\label{eq37}
\vec{u}_{l}(\vec{r},t)=\frac{1}{(2\pi)^{3/2}}\int_{\mathbf{R}^{3}}\ud\omega\ud^{2}\vec{q}\;\hat{\vec{v}}(\vec{q},\omega)e^{\alpha(\vec{q})z}e^{i(\vec{q}\cdot\vec{R}-\omega t)},
\end{equation}
where $\vec{R}=(x,y)$, $\vec{q}=(q_{x},q_{y})$ is the 2D wave vector and $\alpha(\vec{q})$ is the inverse of the penetration length. The longitudinal mode satisfies the condition $\nabla\times\vec{u}_{l}=\vec{0}$, which in the Fourier space turns into the set of equations
\begin{eqnarray}
\label{eq38}
iq_{y}\hat{v}_{z}-\alpha\hat{v}_{y}&=&0,\\
iq_{x}\hat{v}_{z}-\alpha\hat{v}_{x}&=&0,\nonumber\\
q_{x}\hat{v}_{y}-q_{y}\hat{v}_{x}&=&0.\nonumber
\end{eqnarray}
With account of all these functional dependences, the boundary conditions in the Fourier space become
\begin{align}
\label{eq39}
&\frac{1}{(2\pi)^{1/2}}\int_{\mathbf{R}}\ud k_{z}\left[(i k_{z})\hat{u}_{x,t}+(i q_{x})\hat{u}_{z,t}\right]+2\alpha\hat{v}_{x}=0,\\
&\frac{1}{(2\pi)^{1/2}}\int_{\mathbf{R}}\ud k_{z}\left[(i k_{z})\hat{u}_{y,t}+(i q_{y})\hat{u}_{z,t}\right]+2\alpha\hat{v}_{y}=0,\nonumber\\
&\int_{\mathbf{R}}\frac{\alpha\ud k_{z}}{(2\pi)^{1/2}}\left[(i k_{z})\hat{u}_{z,t}+(1-2\lambda^{2})\left\{(i q_{x})\hat{u}_{x,t}+(i q_{y})\hat{u}_{y,t}\right\}\right]\nonumber\\
&\qquad+[\alpha^{2}+(2\lambda^{2}-1)q^{2}]\hat{v}_{z}=0,\nonumber
\end{align}
where $\lambda=c_{t}/c_{l}$ is a parameter of the solid. Computation of the integrals leads to the following results:
\begin{align}
\label{eq40}
&\frac{1}{(2\pi)^{1/2}}\int_{\mathbf{R}}\ud k_{z}\;\hat{u}_{x,t}=\frac{1}{(2\pi)^{1/2}}\int_{\mathbf{R}}\ud k_{z}\;\hat{u}_{y,t}=0,\\
&\frac{1}{(2\pi)^{1/2}}\int_{\mathbf{R}}\ud k_{z}\;\hat{u}_{z,t}=\frac{-\hbar}{8\pi\rho c_{t}^{2}}\frac{i q_{x} a+iq_{y} b}{\alpha_{t}(q)},\nonumber\\
&\frac{1}{(2\pi)^{1/2}}\int_{\mathbf{R}}\ud k_{z}\;(i k_{z})\hat{u}_{x,t}=\frac{\hbar a}{8\pi\rho c_{t}^{2}}\left[\alpha_{t}-\frac{2k_{1D}}{\pi}\right],\nonumber\\
&\frac{1}{(2\pi)^{1/2}}\int_{\mathbf{R}}\ud k_{z}\;(i k_{z})\hat{u}_{y,t}=\frac{\hbar b}{8\pi\rho c_{t}^{2}}\left[\alpha_{t}-\frac{2k_{1D}}{\pi}\right],\nonumber\\
&\frac{1}{(2\pi)^{1/2}}\int_{\mathbf{R}}\ud k_{z}\;(i k_{z})\hat{u}_{z,t}=0,\nonumber
\end{align}
where $\alpha_{t}(\vec{q})=\left[q^{2}-(\omega/c_{t})^{2}\right]^{1/2}$ and $k_{1D}=2\pi/L$ is the cut-off wave vector along the $Z$ axis. With account of these expressions, the third equation of the set \eqref{eq39} becomes $[\alpha^{2}+(2\lambda^{2}-1)q^{2}]\hat{v}_{z}=0$, which leads to the following dependence for the inverse of the penetration length:
\begin{equation}
\label{eq41}
\alpha(\vec{q})=\left[1-2\lambda^{2}\right]^{1/2}q.
\end{equation}
Combining the first two equations of \eqref{eq39} with account of Eqs. \eqref{eq38} yields the identity
\begin{align}
\hat{v}_{z}(\vec{q},\omega)&=\frac{1}{2q^{2}}\Bigg[\frac{-q^{2}}{\sqrt{2\pi}}\int_{\mathbf{R}}\ud k_{z}\;\hat{u}_{z,t}+\frac{i q_{y}}{\sqrt{2\pi}}\int_{\mathbf{R}}\ud k_{z}(i k_{z})\hat{u}_{y,t}\nonumber\\
&\hspace{1.5cm}+\frac{i q_{x}}{\sqrt{2\pi}}\int_{\mathbf{R}}\ud k_{z}(i k_{z})\hat{u}_{x,t}\Bigg]\nonumber\\
&=\frac{\hbar}{16\pi\rho c_{t}^{2}}\left[\alpha_{t}-\frac{2k_{1D}}{\pi}+\frac{q^{2}}{\alpha_{t}}\right]\frac{iq_{x}a+iq_{y}b}{q^{2}},
\end{align}
where in the last step we have used Eqs. \eqref{eq40}. The functional dependence of $v_{z}$ on the surface can be obtained by inverse Fourier transformation:
\begin{align}
\label{eq42}
&v_{z}(\vec{R},z=0,t)=\frac{1}{(2\pi)^{3/2}}\int_{\mathbf{R}^{3}}\ud\omega\ud^{2}\vec{q}\;\hat{v}_{z}(\vec{q},\omega)e^{i(\vec{q}\cdot\vec{R}-\omega t)}\\
&=\frac{i\hbar}{8(2\pi)^{5/2}\rho c_{t}^{2}}\int_{\mathbf{R}}\ud\omega\,e^{-i\omega t}a\int_{\mathbf{R}^{2}}\ud^{2}\vec{q}\;m(\vec{q},\omega)\frac{q_{x}}{q^{2}}e^{i\vec{q}\cdot\vec{R}}\nonumber\\
&+\frac{i\hbar}{8(2\pi)^{5/2}\rho c_{t}^{2}}\int_{\mathbf{R}}\ud\omega\,e^{-i\omega t}b\int_{\mathbf{R}^{2}}\ud^{2}\vec{q}\;m(\vec{q},\omega)\frac{q_{y}}{q^{2}}e^{i\vec{q}\cdot\vec{R}}\nonumber
\end{align}
where $\ds m(\vec{q},\omega)=\alpha_{t}(q)-\frac{2k_{1D}}{\pi}+\frac{q^{2}}{\alpha_{t}(q)}$. With account of the Jacobi-Anger expansion
\begin{equation}
\label{eq43}
e^{i\vec{q}\cdot\vec{R}}=e^{iqR\cos(\theta-\theta_{R})}=\sum_{n=0}^{\infty}\epsilon_{n}i^{n}J_{n}(qR)\cos\left[n(\theta-\theta_{R})\right],
\end{equation}
where $\epsilon_{0}=1$, $\epsilon_{n}=2,\;\forall n\geq1$, $J_{n}(x)$ is the Bessel function of order $n$ and $\theta_{R}$ is the angle between $\vec{R}$ and the $X$ axis, and with account of the orthogonality of the Fourier basis $\{1\}\cup\{\sin(n\theta),\cos(n\theta)\}_{n\geq1}$, we have that
\begin{align}
\int_{\mathbf{R}^{2}}\ud^{2}\vec{q}\;m(\vec{q},\omega)\left\{\begin{array}{c}
  q_{x}\\
  q_{y}\\
  \end{array}\right\}\frac{e^{i\vec{q}\cdot\vec{R}}}{q^{2}}=
  4\pi i G[R,\omega]\left\{\begin{array}{c}
  \cos\theta_{R}\\
  \sin\theta_{R}\\
  \end{array}\right\},
\end{align}
where $\ds G[R,\omega]=\frac{1}{2}\int_{0}^{q_{2D}}\ud q\; m(\vec{q},\omega) J_{1}(qR)$ and $q_{2D}=\sqrt{2}(2\pi)/L$ is the corresponding two-dimensional cut-off wave vector. Therefore, Eq. \eqref{eq42} becomes
\begin{align}
\label{eq44}
v_{z}(\vec{R},z&=0,t)=\frac{\gamma\hbar \Sigma_{z,0}\mathcal{A}}{8\pi\rho c_{t}^{2}}\left(\frac{2D\Sigma_{z,0}}{\hbar}\right)\times\\
&\Bigg[\left(\frac{x}{R}C[\xi,t]+\frac{y}{R}D[\xi,t]\right)\textrm{Re}\,G[R,\xi]+\nonumber\\
&\hspace{0.5cm}\left(\frac{x}{R}A[\xi,t]+\frac{y}{R}B[\xi,t]\right)\textrm{Im}\,G[R,\xi]\Bigg],\nonumber
\end{align}
where 
\begin{align}
\label{eq45}
C[\xi,t]&=\textrm{Re}\left[\frac{i\xi\eta-\omega_{\textrm{FM}}}{K(\xi)}e^{-i\xi t}\right],\\
D[\xi,t]&=\textrm{Re}\left[\frac{i\xi}{K(\xi)}e^{-i\xi t}\right].\nonumber
\end{align}
At resonance, these functions become
\begin{align}
\label{eq46}
C[\xi\simeq\omega_{\textrm{FM}},t]&=\frac{1}{2\omega_{\textrm{FM}}\eta}[\sin(\omega_{\textrm{FM}}t)+\eta\cos(\omega_{\textrm{FM}}t)],\nonumber\\
D[\xi\simeq\omega_{\textrm{FM}},t]&=\frac{1}{2\omega_{\textrm{FM}}\eta}\cos(\omega_{\textrm{FM}}t).
\end{align}

\section{Discussion}\label{Discussion}

We have computed acoustic waves generated by the spin precession. Our results are parameter-free in a sense that they do not contain any unknown parameters of spin-phonon interaction. The computed amplitudes of the acoustic waves have been expressed in terms of parameters that can be independently obtained from macroscopic experiments, such as FMR frequency, speed of sound, etc. The effect comes from the elastic twists generated by torque associated with the spin precession. It is a microscopic counterpart of the Einstein - de Haas effect. Its observation would provide an interesting test of theoretical concepts. 

Generation of these acoustic waves by a precessional nanoparticle can be detected experimentally by means of the TEM set-up. The current spatial resolution is about 0.01 \AA. Potential candidates to host an observable acoustic deformation of the substrate should combine high (uniaxial) anisotropic nanoparticles and a soft substrate. In the following, we focus on the case of an FePt nanoparticle deposited on a gold substrate.

As an elastic medium, gold is characterized by the values $\rho=19.3$ g/cm$^{3}$ for the mass density and $c_{t}=120000$ cm/s for the transversal speed of sound. As a consequence of the crystalline ordering of the face-centered tetragonal structure, FePt nanoparticles in the $L1_{0}$ phase show a very high uniaxial magnetocrystalline anisotropy. The corresponding lattice constants are $a=3.85$ {\AA} and $c=3.71$ {\AA}. Within a tetragonal cell there are 4 Pt atoms (at the center of the lateral faces) and 10 Fe atoms (2 at the center of the top and bottom faces and 8 at the vertexes of the cell). With account of the spin $S=5/2$ for these atoms, the spin density of the L1$_{0}$-phase FePt nanoparticles is $\rho_{S}^{\textrm{FePt}}=1.82\cdot10^{23}$ cm$^{-3}$. The ferromagnetic resonance frequency of these nanoparticles is $\omega_{\textrm{FM}}=5.63\cdot10^{10}$ Hz. Therefore, assuming a cubic geometry of the spin cluster we have $\Sigma_{z,0}=\rho_{S}^{\textrm{FePt}}\cdot L^{3}$, where $L$ is the size of the nanoparticle. 

\begin{figure}[htbp!]
\center
\includegraphics[scale=0.28]{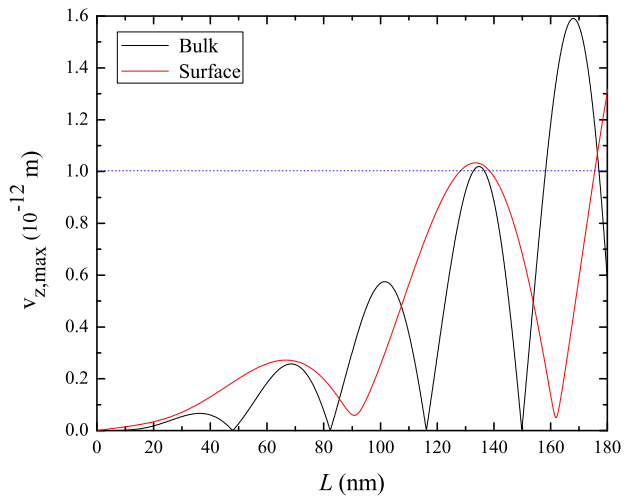}
\caption{(Color online) Functional dependence of the maximum amplitude of both bulk (black) and surface (red) acoustic waves on the size $L$ of the nanoparticle when measured at the distance $R=2L$. The blue dotted line represents the resolution $0.01$ {\AA} of the STM technique.}
\label{Fig1}
\end{figure}
As a criterion to determine whether the effect is measurable we estimate the maximum amplitude of the induced acoustic wave at the distance from the particle that equals its radius and compare it to spatial resolution of the STM technique. This basically suggests that there will be a measurable effect in the immediate vicinity of the particle within the nanoscale range. Fig. 1 shows the dependences of the maximum amplitude of both bulk and surface contributions to the substrate elastic deformation on the size of the nanoparticle when measured at the distance $R=2L$,
\begin{align}
&|v_{z,\textrm{max}}^{\textrm{Bulk}}|=\frac{\gamma\hbar \Sigma_{z,0}\mathcal{A}}{16\pi\eta\rho c_{t}^{2}}\left(\frac{\omega_{\textrm{FM}}}{c_{t}}\right)^{2}\left|j_{1}[(\omega_{\textrm{FM}}/c_{t})2L]\right|,\label{eq47}\\
&|v_{z,\textrm{max}}^{\textrm{Surface}}|=\frac{\gamma\hbar \Sigma_{z,0}\mathcal{A}}{16\pi\eta\rho c_{t}^{2}}\sqrt{\left[\textrm{Re}\,G[2L,\omega_{\textrm{FM}}]\right]^{2}+\left[\textrm{Im}\,G[2L,\omega_{\textrm{FM}}]\right]^{2}},\nonumber
\end{align}
where we have approximated $\ds \frac{2D\Sigma_{z,0}}{\hbar}\simeq\omega_{\textrm{FM}}$ ($H_{0}\lesssim100$ Oe) and considered $\eta=0.001$. Notice that for sizes in the ballpark of $L=135$ nm both bulk and surface acoustic waves show a maximum amplitude bigger than the spatial resolution of the STM technique, so that these elastic deformations could be experimentally detected. The detection of only induced surface acoustic wave could be performed by either considering bigger sizes (in the ballpark of $L=180$ nm, where the maximum amplitude of the bulk wave is strongly suppressed) or using nanoparticles with smaller ferromagnetic resonance frequency. The latter is due to the explicit quadratic dependence of Eq. \eqref{eq47} on this frequency whereas the maximum amplitude of the surface contribution is less sensitive to it.

Expression \eqref{zeta} can be approximated by $\ds \zeta(\omega_{\textrm{FM}})\simeq\frac{\hbar\omega_{\textrm{FM}}\Sigma_{z,0}\Xi(\omega_{\textrm{FM}})}{3(2\pi)^{2}\rho c_{t}^{2}}$, so that for FePt nanoparticles we have $|\zeta|\leq7\cdot10^{-4}$ within the range of sizes considered above. Therefore, the results obtained are consistent with the assumption $|\zeta|\ll1$ made in Sec. 3.
Note that particles detectable by means of TEM via acoustic waves generated by the spin precession would be much smaller than samples required for the detection of the ESR signal in conventional macroscopic measurements \cite{Farle}. 

\section{Acknowledgements}
This work has been supported by the U.S. National Science Foundation through grant No. DMR-1161571.

\end{document}